\pdfoutput=1
\documentclass[manuscript]{aa}
\usepackage{graphicx}
\usepackage{txfonts}

\begin{document}

\title{Propagation into the heliosheath of a large-scale solar wind disturbance bounded by a pair of shocks.}

\author{
           E. Provornikova \inst{1} \inst {3}
          \and
           M. Opher \inst{2}
           \and
           V. Izmodenov \inst{3} \inst{4} \inst{5}
           \and 
           G. Toth \inst{6}
          }

   \institute{ Boston University, 725 Commonwealth Ave, Boston MA 02215, USA\\
                  \email{eprovorn@bu.edu}           
         \and
         Boston University, 725 Commonwealth Ave, Boston MA 02215, USA\\
         \email{mopher@bu.edu}
         \and
          Space Research Institute(IKI), Russian Academy of Sciences, 84/32 Profsoyuznaya St, Moscow 117997, Russia
          \and
             Lomonosov Moscow State University, Faculty of Mechanics and Mathematics, Moscow 119991, Russia
          \and
             Institute for Problems in Mechanics, Russian Academy of Sciences, 101-1 Prospect Vernadskogo, Moscow 119526, Russia\\
             \email{izmod@ipmnet.ru}
           \and
             University of Michigan,2455 Hayward, Ann Arbor MI 48109, USA\\
             \email{gtoth@umich.edu}      }

   \date{Received  date; accepted  date}

\abstract 
{
After the termination shock (TS) crossing, the Voyager 2 spacecraft has been observing strong variations of the magnetic field and solar wind parameters in the heliosheath. Anomalous cosmic rays, electrons, and galactic cosmic rays present strong intensity fluctuations. Several works suggested that the fluctuations might be attributed to spatial variations within the heliosheath. Additionally, the variability of the solar wind  in this region is caused by different temporal events that occur near the Sun and propagate to the outer heliosphere. 
} 
{
To understand the spatial and temporal effects in the heliosheath, it is important to study these effects separately. In this work we explore the role of shocks as one type of temporal effects in the dynamics of the heliosheath. Although currently plasma in the heliosheath is dominated by solar minima conditions, with increasing solar cycle shocks associated with transients will play an important role.
} 
{
We used a 3D MHD multi-fluid model of the interaction between the solar wind and the local interstellar medium to study the propagation of a pair of forward-reverse shocks in the supersonic solar wind, interaction with the TS, and propagation to the heliosheath.
} 
{
We found that in the supersonic solar wind the interaction region between the shocks expands, the shocks weaken and decelerate. The fluctuation amplitudes of the plasma parameters vary with heliocentric distance. The interaction of the pair of shocks with the TS creates a variety of new waves and discontinuities in the heliosheath, which produce a highly variable solar wind flow. The collision of the forward shock with the heliopause causes a reflection of fast magnetosonic waves inside the heliosheath.
} 
{}


\titlerunning{Propagation of a shock pair into the heliosheath}
\authorrunning{E. Provornikova et al.} 

\maketitle

%

\section{Introduction}

The heliosheath, a region of the shocked solar wind between the termination shock (TS) and the heliopause (HP), has been explored by Voyager 1 and 2 after the TS crossing. Voyager 1 crossed the TS at a distance of 94 AU from the Sun in December 2004 \citep{stone05} and Voyager 2 crossed it at 84 AU in August 2007 \citep{rich08}. The Voyager spacecraft revealed an important property of the outer heliosphere structure - the asymmetry of the TS - and have been providing the first measurements of the heliosheath. Strong fluctuations of the magnetic field, the solar wind density, the velocity, and the temperature were observed by Voyager 2 in the heliosheath during about one year after the TS crossing \citep{rich11,bur10}. Fluctuation amplitudes have been decreasing as Voyager 2 moved deeper into the sheath. \citet{rich09} also reported quasi-periodic variations of the solar wind velocity components and flow angles with a period of 110 days observed by Voyager 2 in the heliosheath. The plasma experiment is not working on Voyager 1, but data from the magnetic field experiment show large-scale magnetic field fluctuations in the heliosheath \citep{bur2006, burne10}. Observations from the Voyager spacecraft show that the heliosheath is a region with a highly variable and complex plasma flow. 

Different temporal and spatial effects may cause these dynamic flows in the heliosheath. It is important therefore to investigate the effects of temporal solar wind structures on the heliosheath to separate spatial and temporal plasma variations. Several solar wind phenomena contribute to temporal variations: the 11-year solar cycle variations, large-scale structures such as interplanetary coronal mass ejections (ICMEs) \citep{rich06,rich05} and merged interaction regions (MIRs) \citep{burmcd93} that usually occur at solar maximum; corotating interaction regions (CIRs) formed at the declining phase of solar activity \citep{bur97,bur03}; and CME- and CIR-driven shocks  \citep{bur94, wang01, wang02, rich06grl}. These structures generate significant changes in the solar wind parameters. Measurements show that during an 11-year solar cycle the solar wind ram pressure changes by factor of 2 from solar minimum to maximum. CIRs are characterized by an enhanced magnetic field, plasma density, and pressure, and are bounded by a pair of shocks. Observations of CIRs by Voyager 2 and Pioneer 10 in the inner heliosphere within 10 AU showed that the ram pressure in CIRs may increase by factor of 15-30 \citep{gaz00}. At larger heliospheric distances,  CIRs expand and merge to form CMIRs \citep{bur83, bur03}. Voyager 2 data during the solar minimum in 1994-1995 near 45 AU show the sequence of recurring sharp, shock-like increases in the solar wind speed that resemble very much forward shocks \citep{laz99}. These structures are associated with CMIRs. Changes of the solar wind parameters were smaller, but the ram pressure changed by factor of 2-4 at the shocks. Voyager 2 observations of solar maximum plasma between 65-70 AU indicated periodic fluctuations with a correlated solar wind speed, density, and magnetic field, which increases the solar wind ram pressure by a factor of 10 within timescales of 6-12 months (\citet{rich03}). Supposedly, these structures are candidates for MIRs, which are formed from merging transient solar wind events. All these solar wind structures propagate to the TS and beyond and affect the solar wind flow in the heliosheath.

Effects of the 11-year solar cycle variations on the interaction of the solar wind with the local interstellar medium (LISM) were studied by many authors \citep{kar95,bz98,sf03a,sf03b,izmal04a,izmal04b,izm05,izm08,zank03,pog09}. Models applied different boundary conditions that simulated changes of the solar wind dynamic pressure during the solar cycle. Some global models that considered the solar cycle effects ignored the interstellar H atoms or used simplified fluid or multi-fluid approximations to describe the neutrals. A kinetic description for interstellar H atoms (which is necessary because of their large mean free path) was employed in \citet{izm05,izm08}. The models showed that the TS and HP oscillate in response to varying solar wind ram pressure in the solar cycle. Using realistic boundary conditions in a time-dependent 2D kinetic-hydrodynamic model, \citet{izm08} obtained that the TS reflects variations of the solar wind dynamic pressure in 1-1.2 year and the TS location fluctuates by $\pm$ 7.5 AU. The model with 11-year periodic boundary conditions by \citet{izm05} showed that the HP varies less - by $\pm$ 2 AU near its mean value. Therefore, the boundaries of the heliosheath are constantly in motion, and the heliosheath is expected to be a highly dynamic region. 

Interaction of the TS with various interplanetary disturbances from upstream was studied by \citet{bar93}, \citet{nb94a, nb94b}, \citet{stein94}, \citet{story95,story97}, \citet{bar96a, bar96b}, \citet{rat96}, \citet{zank03}, \citet{barpu04} and others in 1D and 2D hydrodynamic and MHD models. The models studied the propagation of solar wind shock waves, contact discontinuities, forward-reverse shock pairs, and ram pressure pulses through the TS and predicted the flow structure downstream of the TS. Recent works by \citet{wash07,wash11} explored the effects of realistic pulses of the solar wind ram pressure on the heliosheath flow. They performed a 3D MHD simulation using Voyager 2 plasma data for the boundary conditions upstream of the TS. They showed that when the ram pressure pulse collides with the TS, (1) the TS moves away from the Sun; (2) a large-amplitude magnetosonic wave is generated downstream of the TS; (3) the magnetosonic wave is reflected inside the heliosheath; and (4) the collision of the reflected wave with the TS causes the motion of the TS toward the Sun. 

Effects of CIRs in the heliosphere were modeled by \citet{pizzo94}, \citet{pizzo94a,pizzo94b} in a 3D MHD corotating model of the solar wind flow. \citet{borovikov2012} modeled the evolution of CIRs in the outer heliosphere and in the heliosheath in a 3D time-dependent model of the solar wind interaction with the LISM. Their results show that CIRs create a complex non-stationary plasma flow in the heliosheath and produce entropy and fast magnetosonic waves. \citet{bur03} investigated the evolution of  the magnetic field fluctuations induced by CIRs in a 1D MHD model using realistic solar wind parameters in 1995 during the solar minimum. The model predicted  broad regions with enhanced magnetic field caused by CIRs in the outer heliosphere.

MIRs and global MIRs (GMIRs) are non-periodic large-scale disturbances in the heliosphere compared to CIRs. MIRs are regions with enhancements in the density and magnetic field strength and an increase of the bulk speed \citep{burmcd93, burness1994}. These structures usually evolve from transient events that occur near the Sun, such as CMEs and isolated fast streams \citep{rich06grl, whang01, richardson02}. Interplanetary CMEs and MIRs were frequently observed by the Wind, Ulysses, and Voyager spacecraft. Observations from Ulysses showed that these structures  may drive a pair of forward and reverse shocks \citep{whang01, gosling1994, manchester2006}. Before crossing the TS, Voyager 2 observed a MIR at a distance of 79 AU from the Sun \citep{rich06}. Using a 1D MHD model, \citet{rich06} showed that the MIR was produced by a CME merged with high-speed plasma streams. The model also showed that the MIR is bounded by forward and reverse shocks which agrees with Voyager 2 observations.

In general, the structure of a MIR is very complex because of interaction with ambient plasma and merging with other disturbances in the solar wind. In this work, we aim to explore the dynamical effects of a pair of shocks relevant to MIRs  that propagate to the outer heliosphere and heliosheath. We describe in detail the interaction of a pair of shocks with the TS, propagation in the heliosheath, and the interaction with the HP and explain the formation of MHD discontinuities and waves in the heliosheath that create the highly variable flow. We use a 3D MHD global model of the interaction between the solar wind and the LISM \citep{op09}.

Previously, interaction of a pair of forward-reverse shocks with the TS was studied by \citet{story95, story97} in the frame of 1D planar gas-dynamic and MHD models. \citet{bar96a, bar96b}, \citet{barpu04} considered 2D MHD interaction of the TS with forward and reverse interplanetary shocks. They showed that the collision of shocks with the TS generates a large number of wave modes and shocks downstream of the TS. We compare our results with the previous models and extend the study to the interaction of shocks with the HP.

The outline of the paper is the following. In section 2 we briefly describe the model. In section 3 we present the results: (1) the evolution of a pair of forward-reverse shocks in the supersonic solar wind, (2) the interaction with the TS, and (3) the propagation through the heliosheath and interaction with the HP. Conclusions are given in section 4.


\section{Description of the model}

To model the propagation of the solar wind disturbances to the heliosheath we used a global 3D MHD model of the solar wind interaction with the LISM \citep{op09}. The model is based on the BATS-R-US code \citep{toth12}. The model includes the influence of interstellar hydrogen atoms that penetrate into the heliosphere. Plasma protons interact with H atoms in a charge-exchange process. Because of the large mean free path of H atoms compared to the size of the heliospheric interface, a kinetic approach is required for the neutral component \citep{izm00,izm01}. Several global models use a multi-fluid approximation to describe H atoms \citep{zank03, op09} due to the computational difficulty of solving a kinetic equation. The multi-fluid model reproduces the global structure of the outer heliosphere well, but it has some limitations \citep{alizm05}. The comparison of the multi-fluid and kinetic models performed by  \citet{alizm05} showed that solutions for the plasma differ by $5 \%$. The difference in the parameters of the interstellar hydrogen atoms between the kinetic and multi-fluid models is significant. In the present work, we are interested in the plasma flow and used the multi-fluid approximation for the description of H atoms. A study of non-stationary flows in the heliosheath based on a 3D MHD model with a kinetic treatment of H atoms (as was done by \citet{bibi}) will be performed in the future.

Separate systems of Euler equations with source terms for charge exchange were solved for each of the four H atom populations, and ideal MHD equations with source terms were solved for the plasma component (for more details, see \citet{op09}). The solar and interstellar magnetic fields are included in the model. The inner boundary of a domain is a sphere at 30 AU and the outer boundary is at $x= \pm 1000 AU, y= \pm 1000 AU, z= \pm 1000 AU$. Parameters of the solar wind at the inner boundary were chosen to match the values obtained by \citet{izm09} at 30 AU: $V_{sw}=417 km/s$ , $n_{sw}=8.74 \times 10^{-3} cm ^{-3}$, $T_{sw}=1.087 \times 10^5 K$ and the Parker spiral magnetic field $B_{sw}= 7.17 \times 10^{-3} nT$  at the equator. Our simulations are different from \citet{op11} in that we assumed that the magnetic axis is aligned with the solar rotation axis.
For a steady-state solution of the interaction between the solar wind and the LISM, the solar wind flow at the inner boundary was assumed to be spherically symmetric. For the interstellar plasma we assumed: $V_{LISM} = 26.4 km/s$, $n_{LISM}= 0.06 cm^{-3}$, and $T_{LISM}=6519 K$. The number density of H atoms in the interstellar medium is $n_{H_{LISM}}=0.18 cm^{-3}$, the velocity and temperature are the same as for the interstellar plasma. The coordinate system is such that the Z-axis is parallel to the solar rotation axis, the X-axis is $5 \degr$ above the direction of interstellar flow and Y completes the right-handed coordinate system (a schematic figure can be found in \citet{bibi}).

We assumed that the interstellar magnetic field has a magnitude $B_{LISM}=4.4 \mu G$ and the orientation is such that the angle between the interstellar flow velocity ${\bf V}_{LISM}$ and ${\bf B}_{LISM}$ is $20 \degr$, and the angle between the $({\bf B}_{LISM}, {\bf V}_{LISM})$ plane and  the solar equator plane is $60 \degr$. This direction for the interstellar magnetic field is close to the hydrogen deflection plane \citep{lal05} and reproduces the TS asymmetry in the Voyager 1 and Voyager 2 directions \citep{op09}.

In this paper, we consider the propagation of the solar wind disturbances along the Voyager 2 trajectory. We designed a numerical grid with a highly refined cone extending from 30 AU beyond the HP along the Voyager 2 trajectory. The entire grid in the domain has $10^7$ cells with nine levels of refinement, ranging from scales of 0.5 AU at the inner boundary and in the cone to 32 AU at the outer boundary. To solve the equations numerically, we used a second-order HLLE scheme with a monotonized central flux limiter function.

\begin{figure}
\centering
\includegraphics[width=\hsize, angle=270]{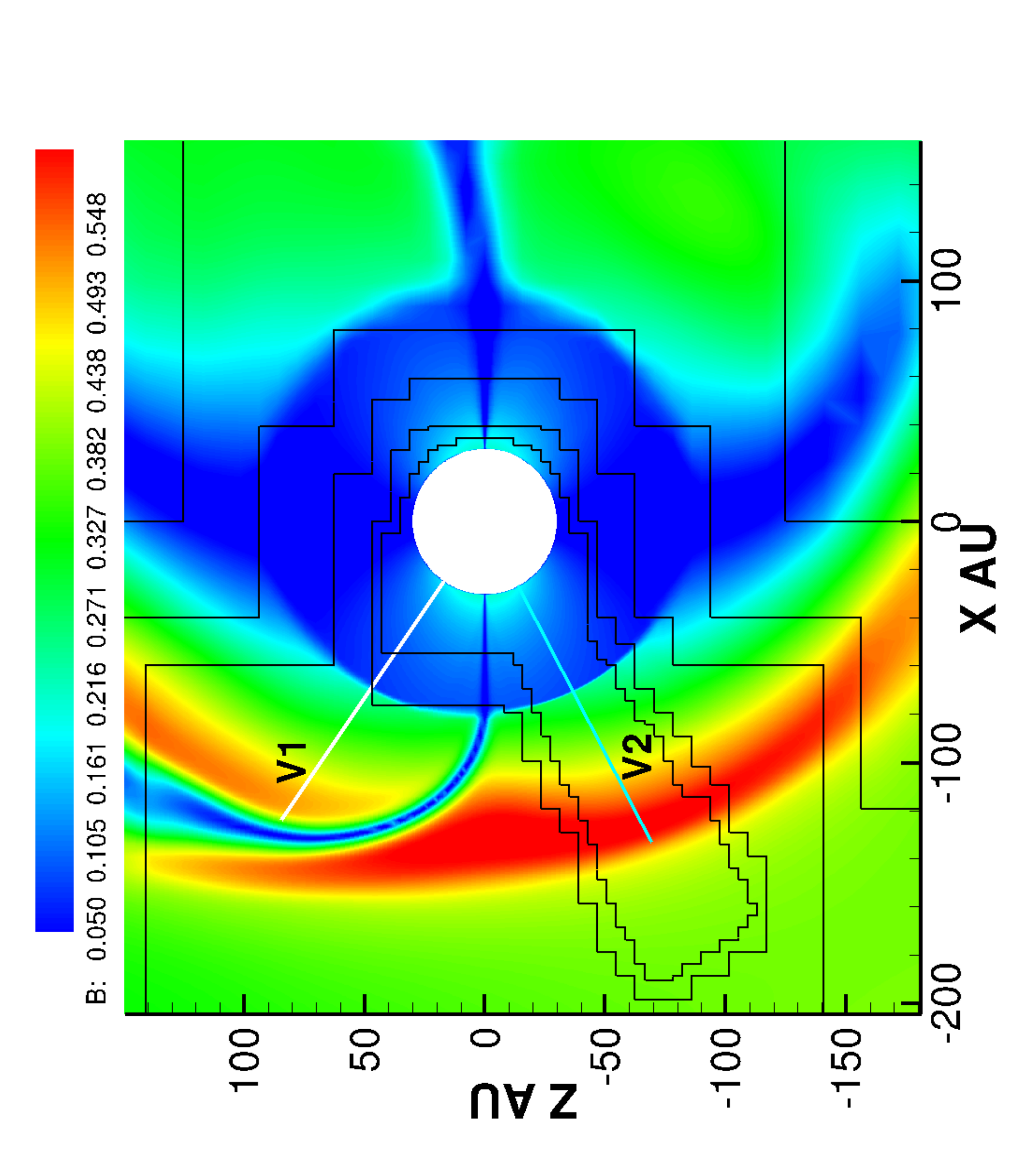}
\caption{Plane cut through the solar rotation axis and Voyager 2 direction (Oz-V2 plane) from a 3D MHD simulation showing the magnitude of the magnetic field (nT). The Voyager 1 trajectory is $\sim 30\degr$ above the solar equator (white line) and that of Voyager 2 is $\sim 30 \degr$ below the equator (blue line). Black boundaries show the regions with different levels of grid refinement, the resolution in the cone along the Voyager 2 trajectory is 0.5 AU. \label{fig1}}
\end{figure}

Fig. \ref{fig1} shows the magnitude of the magnetic field in the  plane through the solar rotation axis and the direction of the Voyager 2 trajectory (hereafter Oz-V2 plane) for the steady-state solution. The inclined ${\mathbf B}_{LISM}$ produces an asymmetry of the heliosphere that pushes the TS and HP closer to the Sun in the south. This stationary solution was used to initiate the disturbances at the inner boundary.
  
To generate a pair of forward and reverse shocks we increased the solar wind speed from $417 km/s$ by factor of a 1.5 at the inner boundary along the Voyager 2 direction. We assumed that the angle between the discontinuity surface and the solar wind velocity vector in the direction of Voyager 2 is 90 degrees (see Fig. 2 below). The angular size of a high-speed stream is $30 \degr$ in $\theta$  and $30 \degr$ in $\phi$ (where $\theta$ and $\phi$ are the latitudinal and azimuthal angles in the HGI coordinate system). 
The shocks and the interaction region between them are located within the highly resolved grid cone during the entire time of the simulation. We split the study of the propagation of the pair of forward-reverse shocks into the following sections: 1) propagation in the supersonic solar wind out to the TS; 2) interaction with the TS; 3) propagation in the heliosheath; and 4) interaction with the HP and the posterior evolution of the reflected waves in the heliosheath.

\section{Results}
\subsection{Evolution of a pair of shocks in the supersonic solar wind} 

\begin{figure}
\centering
\includegraphics[width=\hsize]{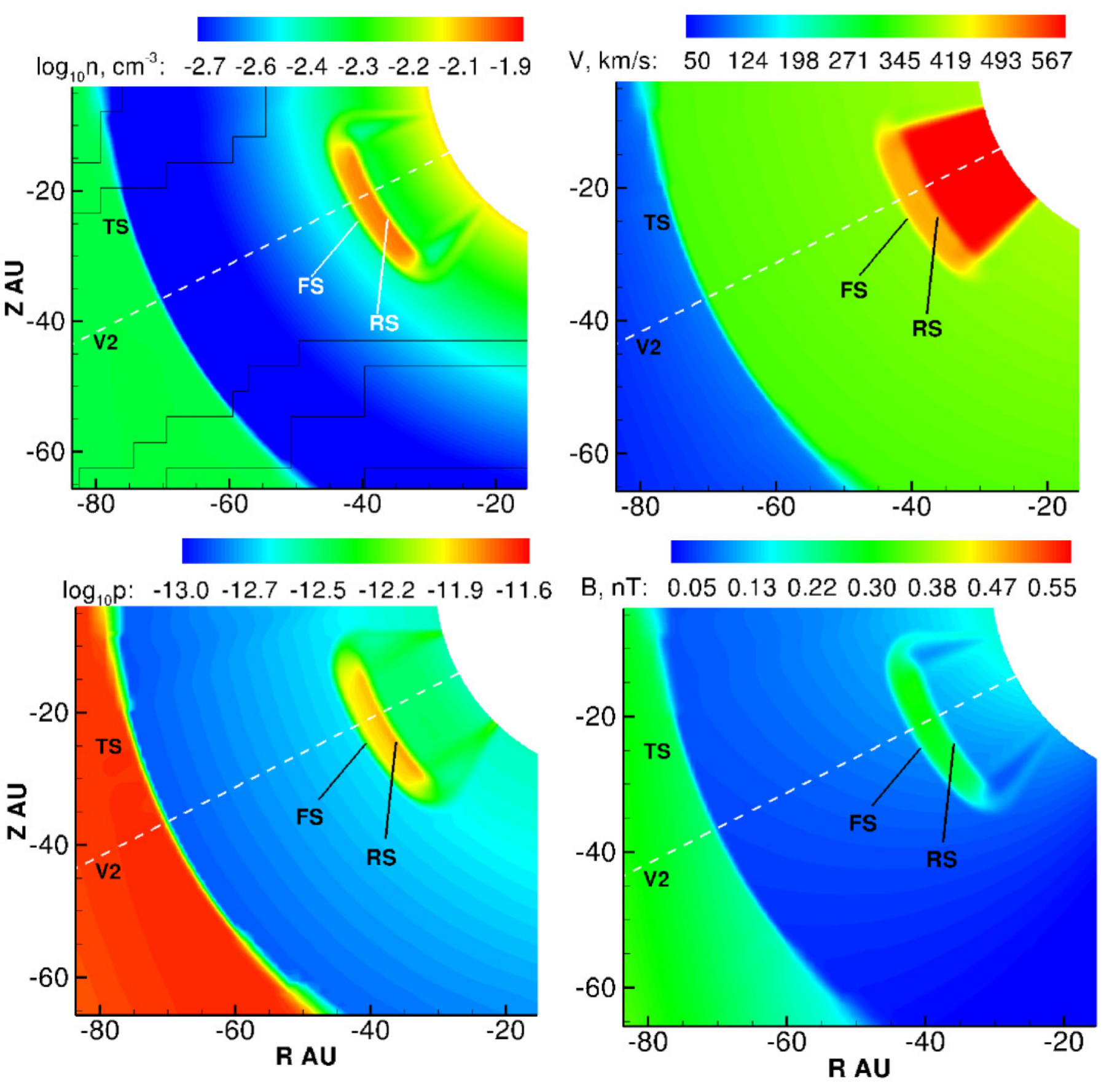}
\caption{Cut from a 3D MHD simulation showing the log of the solar wind number density ($cm^{-3}$), the velocity ($km/s$), the log of thermal pressure ($dyn/cm^2$), and the magnitude of the magnetic field (nT) in the Oz-V2 plane. The plots show the formation of a forward-reverse shock pair that propagates along the Voyager 2 direction (dashed white line). The forward shock (FS), reverse shock (RS), and termination shock (TS) are denoted. Black boundaries show the regions of the grid refinement. \label{fig2}}
\end{figure}

After the initiation of a high-speed stream in the Voyager 2 direction, an arbitrary discontinuity separating the fast and ambient slow solar wind forms at the front. It decays and two shocks - forward and reverse - are formed. Figure \ref{fig2} presents the solar wind number density, the velocity, the thermal pressure and the magnetic field in the Oz-V2 plane at $t=0.3$ yr ($t=0$ corresponds to the steady-state solution at the time that the disturbance was initiated). It can be seen that a large-scale interaction region with enhanced density, thermal pressure, and magnetic field is formed in the supersonic solar wind; the region is bounded by the forward and reverse shocks.  Indeed, a solution of the Riemann problem in MHD shows that a tangential discontinuity (TD) must form between the two shocks. In figure 3 (blue curve) an increase of plasma density and magnetic field and decrease of temperature can be identified between the shocks. These parameter changes correspond to the TD. The numerical scheme used in the simulation diffuses tangential discontinuities. For this reason we focus on the evolution of the two shocks in our MHD analysis of the plasma flow.

At $t=0.3$ yr the interaction region passes the distance 45 AU from the Sun and the compression ratios are $\delta_{FS}=2.4$ and $\delta_{RS}=2.1$ for the forward and reverse shocks, respectively. The total pressure $p_{tot}=p_{th}+p_{B}$ and dynamic pressure $\rho V^2$ increase by factors of 5 and 4, respectively, in the interaction region compared to the undisturbed solar wind upstream of the forward shock.

The characteristics of the shocks and the variations of the solar wind parameters in the interaction region change while they propagate away from the Sun. The shocks weaken due to a change of the background solar wind parameters with heliospheric distance and spherical expansion of the interaction region. Our model shows that the forward shock compression ratio $\delta_{FS}$ decreases by 13 \% from 40 AU out to the TS. For the reverse shock, $\delta_{RS}$ decreases by $\sim 20 \%$.  The forward shock speed decreases by $\sim 60 \%$ from 288 to 120 km/s near the TS (with respect to the solar wind upstream of the forward shock). The reverse shock accelerates in the solar wind frame, the shock speed increases from 120 km/s to 245 km/s. 

The interaction region expands radially with time since the forward and reverse shocks move in opposite directions in the solar wind frame. The width increases about three times from 4.5 AU at a distance $\sim 40 $ AU from the Sun to 13.5 AU near the TS. 

\begin{figure}
\centering
\includegraphics[width=\hsize]{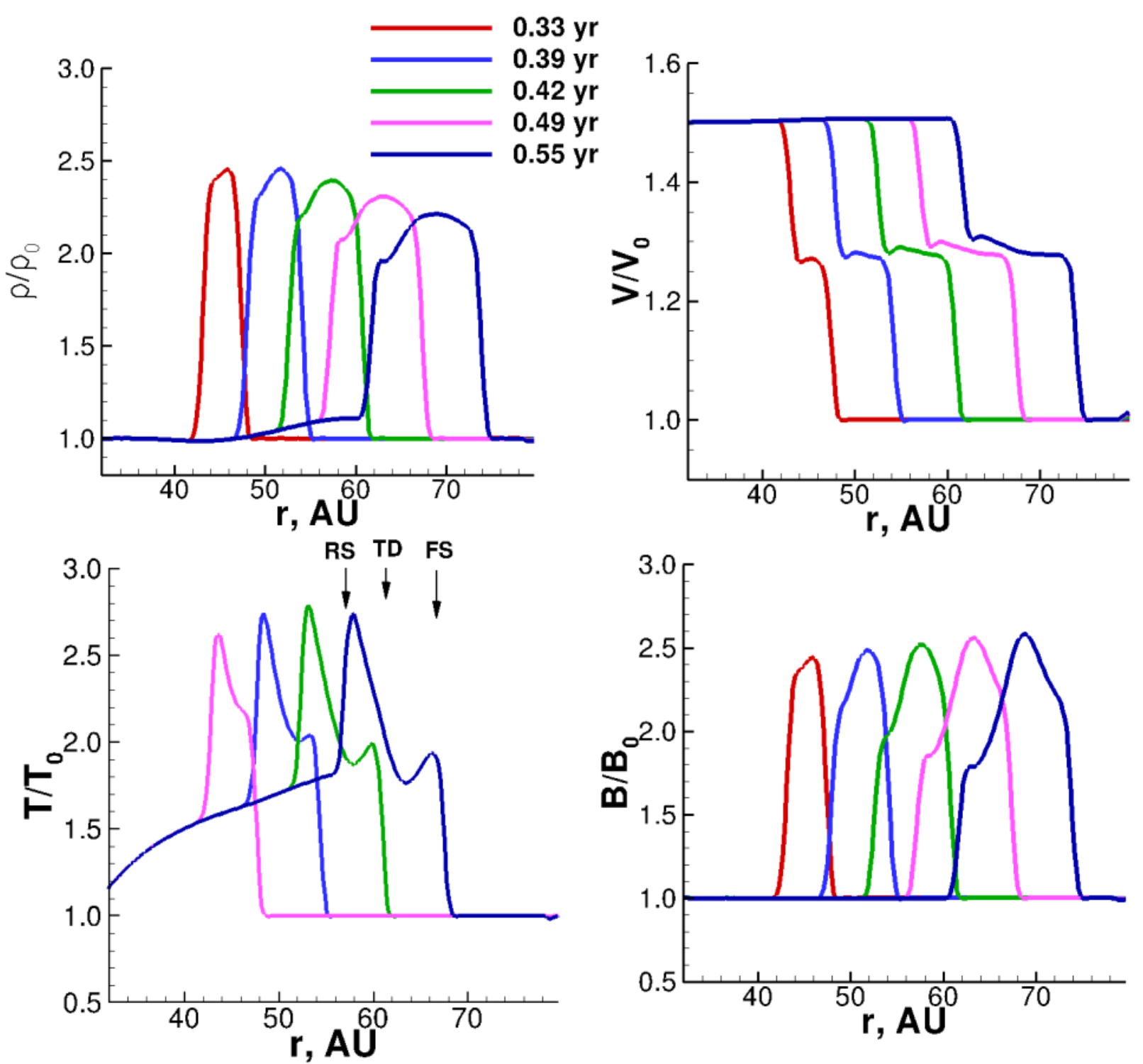}	  
\caption{Profiles of normalized density $\rho/\rho_0$,  velocity $V/V_0$, temperature $T/T_0$,  and  magnetic field $B/B_0$ along the Voyager 2 trajectory for different times showing the change in amplitudes of the fluctuations in the interaction region. Here $\rho_0$, $V_0$, $T_0$ $B_0$ denote the steady-state solution. Notations: FS - forward shock, TD is the tangential discontinuity, RS - reverse shock. \label{fig3}}
\end{figure}

To explore the 3D effects of the flow, we analyzed the variation changes of the solar wind parameters in the interaction region while it propagates in the supersonic solar wind. To reveal the latitudinal dependence we investigated the different directions in the V2-Oz plane: $\theta_1=30 \degr$ corresponding to the Voyager 2 direction; $\theta_2 = 20 \degr$; $\theta_3= 40 \degr$ (here the angle is measured from the equatorial plane toward the south). These directions are within the high resolved cone of our computational grid. Figure \ref{fig3} shows the evolution of the interaction region along the Voyager 2 direction.  Plasma parameters vary inside the interaction region because of the TD and possibly other waves. A region of highest density and magnetic field is created through the pile-up of the solar wind plasma behind the forward shock. The interaction region between the shocks exhibits strong variations of the solar wind parameters. At a distance of 45 AU from the Sun, the magnetic filed strength increases by a factor of 2.4 inside the interaction region, the solar wind density by 2.4, and the temperature by 2.5. As the disturbance evolves, the fluctuation amplitude of normalized magnetic field $B/B_0$ increases by $6 \%$, the amplitude of the density  fluctuation $\rho/\rho_0$ decreases by $12 \%$, the temperature fluctuation $T/T_0$ increases non-monotonically by $10 \%$, and the change in the velocity $V/V_0$ is negligibly small. Here $\rho_0$, $V_0$, $T_0$ and $B_0$  refer to the steady-state values. Along the directions  $\theta_2$ and $\theta_3$ the model shows the same fluctuations behavior for $\rho/\rho_0$, $V/V_0$, and $p_{tot}/p_{tot_0}$. However, in the direction $\theta_3$ the amplitude of $B/B_0$ fluctuation increases by $8 \%$, while along the $\theta_2$ it remains nearly constant. The difference in evolution of the magnetic field fluctuations is caused by the variation of the background Parker magnetic field $B_0$ with latitude. Latitudinal variations in evolution of the interaction region show that the 3D effects of the flow induced by realistic large-scale solar wind disturbances could be important.

Our model shows increasing magnetic field fluctuations in the interaction region with distance, which is different from what \citet{bur03} found in their 1D model. They found that the amplitude of $B/B_0$ in CIRs decreases by $\sim 4 \%$. This discrepancy is due to a difference in the boundary conditions between the models  and 3D effects taken into account in our model. 

Modeling the evolution of an interaction region bounded by a pair of shocks from 30 AU out to the TS showed that (1) the shocks weaken; (2) the  interaction region expands radially and decelerates; (3) the variations of the solar wind parameters in the interaction region become weaker, and the variations of the magnetic field increase with the heliospheric distance. Our study of the 3D effects of the solar wind flow due to the propagating pair of shocks showed that magnetic field fluctuations become stronger with increasing latitude from the Voyager 2 direction. 

\subsection{Interaction of a pair of shocks with the TS}

When a pair of forward-reverse shocks propagates to the outer heliosphere, the forward shock eventually encounters the TS. \citet{bar96a} studied MHD interaction of a forward interplanetary shock with the stationary TS in a two-dimensional model. They showed that the solution of the problem is determined by five dimensionless parameters:
$ M_{TS}=V_{TS}/a_0, \,\ M_{IPS}=V_{IPS}/a_0, \,\ \beta=8 \pi p/B^2, \theta$, and $\psi_{TS} $, where $V_{TS}$ is the solar wind speed immediately upstream of the TS, $a_0$ is the sound speed upstream of the TS, $V_{IPS}$ is the interplanetary shock speed (in our case this is the speed of the forward shock), $\beta$ is the ratio of  thermal and magnetic pressure,  $\theta$ is the angle between the normals to the TS and the interplanetary shock fronts, and $\psi_{TS}$  is the inclination angle of the interplanetary magnetic field vector relative to the TS front.

In our simulation, $\theta \sim 0$, $\psi_{TS} \sim 0$, $M_{TS} =3.8$, $M_{IPS}=3.5$, $\beta=4.6$. For these values, \citet{bar96a} reported a configuration of the interaction of the forward shock and the TS that was expressed by the following scheme:
$$
TS \rightarrow \leftarrow FS \Rightarrow
  \leftarrow FS' \leftarrow R \,\ C \,\ SS \rightarrow
 TS' \rightarrow,
$$
where the arrows show the directions of motion of the discontinuities after the shocks interact in the solar wind frame. Here $FS'$ is the new modified forward shock, propagating into the heliosheath, $R$ is a slow rarefaction wave, $C$ is a contact discontinuity, $SS$ is a reverse slow shock, and $TS'$ is the modified TS. 

\begin{figure}
\centering
\includegraphics[width=\hsize]{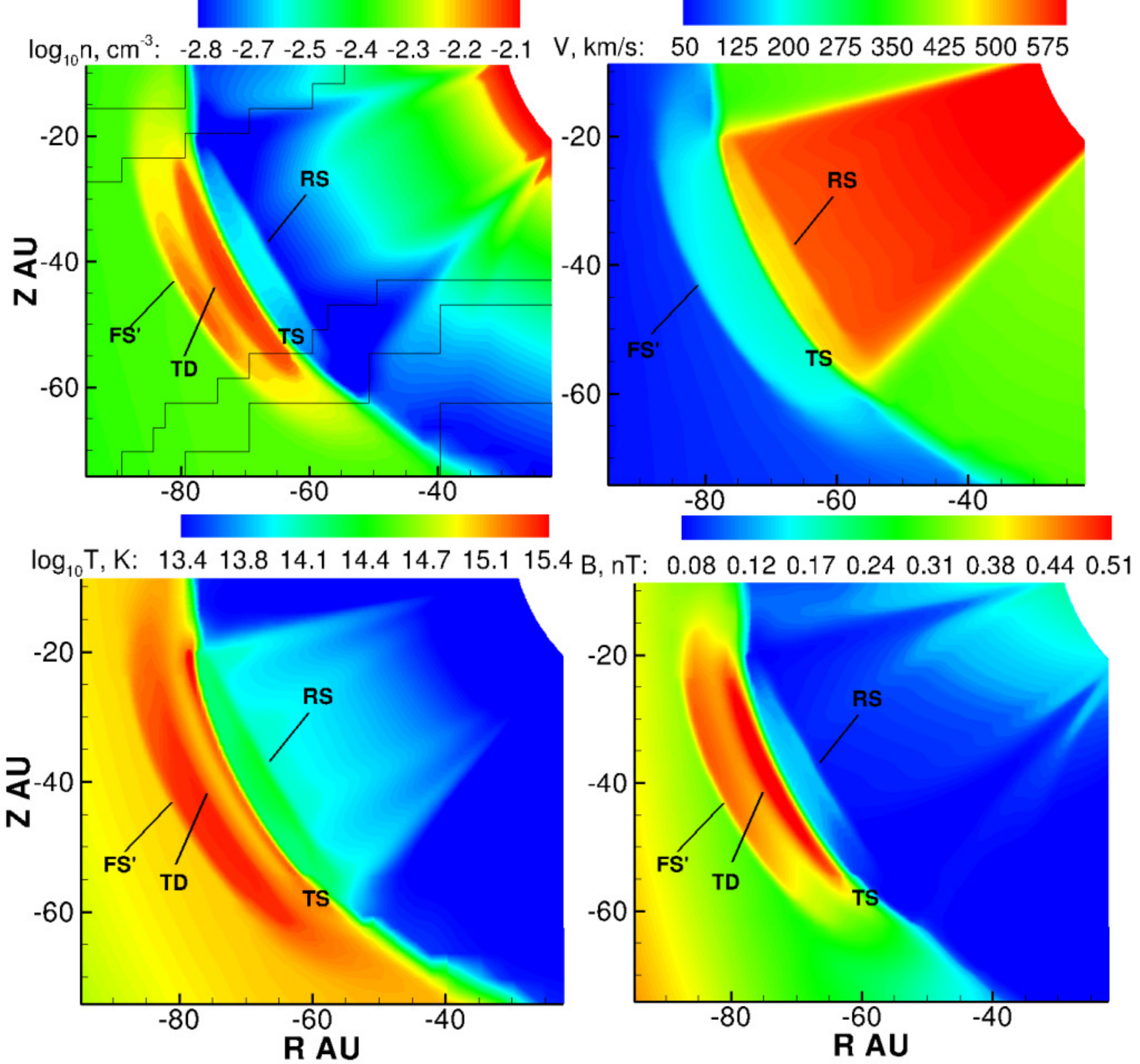}
\caption{Meridional cuts through the Oz-V2 plane from a 3D MHD simulation showing the log of the solar wind number density ($cm^{-3}$), the velocity ($km/s$), the log of temperature ($K$), and the magnetic field magnitude ($nT$) at $t=0.7$ yr  when the forward shock crossed the heliospheric termination shock. FS' denotes the forward shock propagating into the heliosheath, TD is a tangential discontinuity, TS is the termination shock, and RS is a reverse shock upstream of the TS.\label{fig4}}
\end{figure}

Our numerical solution shows structures in the flow similar to the analytical solution of \citet{bar96a}. Figure \ref{fig4} shows the number density, the velocity, the temperature, and the magnetic field in the V2-Oz plane at the moment $t=0.7 yr$ after the forward shock crossed the TS and before the reverse shock encounters the TS. Our results show the formation of a new forward shock (FS') that propagates into the heliosheath,  the modified TS (TS'), and a structure between them characterized by an increasing plasma density and magnetic field intensity accompanied with a decreasing temperature (and thermal pressure) and no change in velocity.  At the same time, the total pressure is constant across this structure, indicating that this is a tangential discontinuity (denoted as TD in Fig. 4).

 \citet{sam06} considered an interaction of an interplanetary shock with the Earth's bow shock. Their solution shows the same qualitative behavior of the plasma parameters between the two shocks after their interaction. The authors suggested that this structure is  a combination of a slow rarefaction wave, a contact discontinuity, and a slow shock. The increasing magnetic field with the decreasing density is generated by the reverse slow shock. This decreasing density is compensated by a stronger increase of density across the contact discontinuity. Velocity variations are negligible. \citet{sam06} pointed out that the increased grid resolution in their model is not sufficient to reproduce the separated discontinuities because of similar velocities of the discontinuities and small changes of MHD parameters. 
 
 Even with the second-order numerical scheme and the spatial resolution of 0.5 AU, our simulation is unable to resolve such small structures. If the slow shock exists in our case according to the solution of  \citet{bar96a}, it may not be resolved by our numerical method and the grid used. High-resolution runs were performed by  \citet{op11}, but they are computationally extremely costly. 

Since the TS is a reverse shock in the solar wind frame of reference, its interaction with the incident forward shock results in a weakening of both shocks. In our case the strength of the forward shock decreased by $\sim 30 \%$ after the TS crossing. The strength of the TS decreased by $7 \%$. Due to the increasing ram pressure behind the forward shock, the TS is displaced by 3 AU away from the Sun.

\begin{figure}
\centering
\includegraphics[width=\hsize]{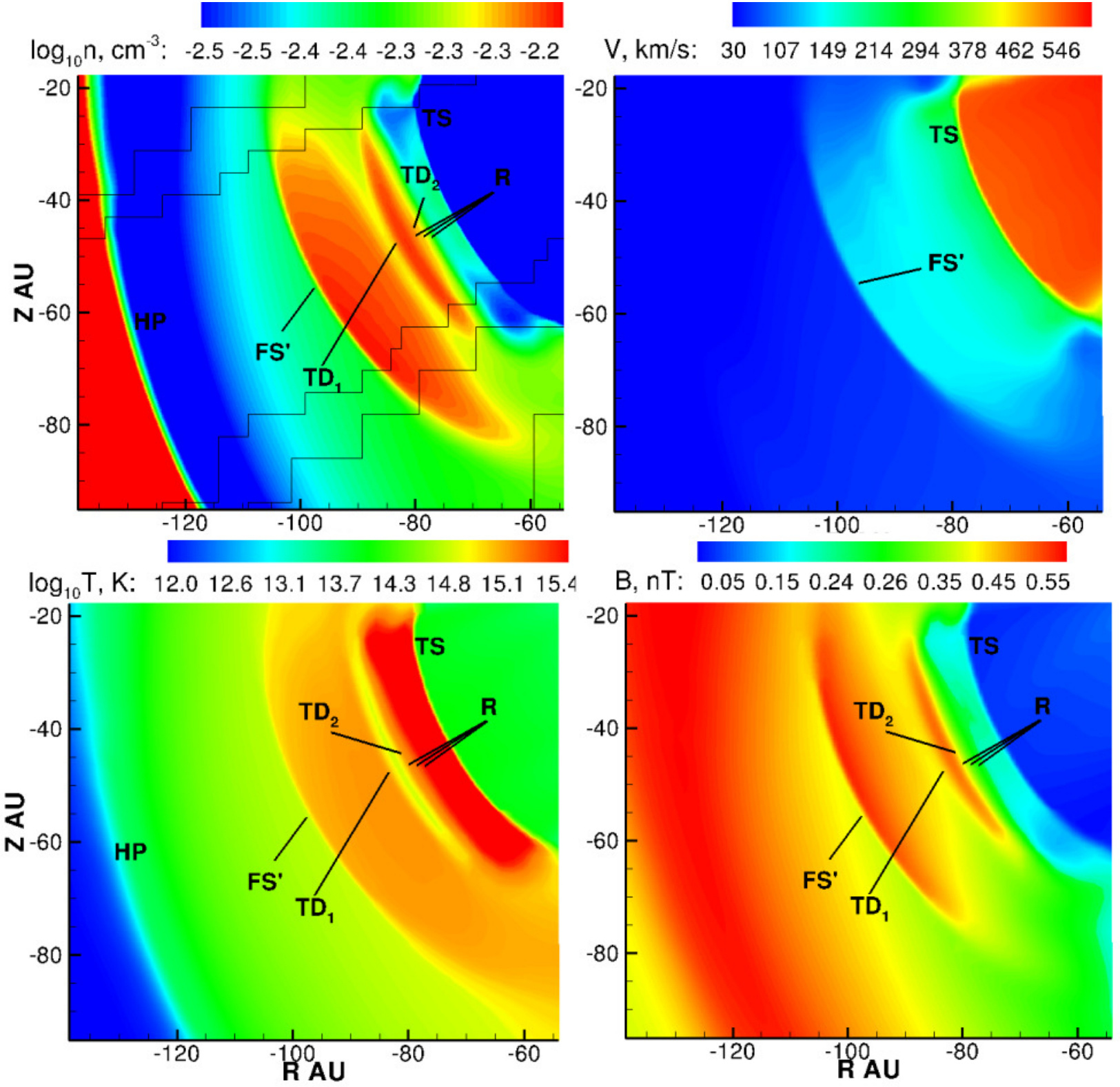}
\caption{Meridional cuts  through the Oz-V2 plane from a 3D MHD simulation showing the log of the solar wind number density ($cm^{-3}$), the velocity ($km/s$), the log of the temperature ($K$), and the magnetic field magnitude ($nT$) at $t=1$ yr after the reverse shock crossed the TS. FS' denotes the forward shock propagating into the heliosheath; $TD_1$ and $TD_2$ are tangential discontinuities; TS is the termination shock; R denotes a rarefaction wave formed after the interaction of the TS and the reverse shock.\label{fig5}}
\end{figure}

	After the passage of forward shock the modified TS' interacts with the reverse shock - the rear side of the interaction region. MHD interaction of the TS with the reverse interplanetary shock was considered by \citet{barpu04} in a two-dimensional model. Their solution gives a configuration consisting of a new TS and several waves propagating downstream of the TS - a fast rarefaction wave, a slow shock, a contact discontinuity, and a slow rarefaction wave. Figure \ref {fig5} shows the results of our simulation; the solar wind parameters are plotted as in Figure \ref{fig4} but at $t=1 yr$ when the reverse shock crossed the TS'. It can be seen that a tangential discontinuity (denoted $TD_2$) and a rarefaction wave R propagate into the heliosheath after the TS' interaction with the reverse shock (Fig. 5). At the tangential discontinuity, the plasma density and the magnetic field decrease and the temperature increases, while the total pressure remain constant. In the rarefaction wave R, the parameters change continuously. The velocity change in the rarefaction wave is negligibly small. Since the configuration of a shock-shock interaction in MHD could be very complex, some additional waves and discontinuities may arise in the heliosheath that are not resolved in the simulation.  Our simulation shows that after merging with the reverse shock, the strength of the TS increases by $10 \%$ and the TS is displaced by 2.6 AU away from the Sun due to increasing solar wind dynamic pressure at the reverse shock.

Therefore, when an interaction region bounded by a pair of shocks passes the TS, a new large-scale region of disturbed solar wind plasma forms in the heliosheath and propagates toward the HP. This new disturbance has a complex structure, it is bounded by the forward shock FS' at the front and the tangential discontinuity $TD_2$ at the rear (see Fig. \ref{fig5}). The TS strength is little changed by the passage of this pair of shocks. The forward shock weakens after interacting with the TS.  In the heliosheath, the forward shock strength varies with latitude: the values of $\delta_{FS}$ obtained within $\pm 10 \degr$ from the Voyager 2 direction showed that $\delta_{FS}$ remains nearly constant toward the higher latitudes and decreases by $5 \%$ to the equator. At the tangential discontinuity $TD_2$, the solar wind number density decreased by $30 \%$. The region between FS' and $TD_2$ is compressed heated plasma with an enhanced density and magnetic field, moving into the heliosheath faster than the ambient plasma (plot of velocity in Fig. \ref{fig5}). The solar wind is highly variable in this region due to formation of waves and discontinuities, as shown above. In the following section, the propagation of this large-scale disturbed region through the heliosheath is discussed.
	
\begin{figure}
\centering
\includegraphics[width=\hsize]{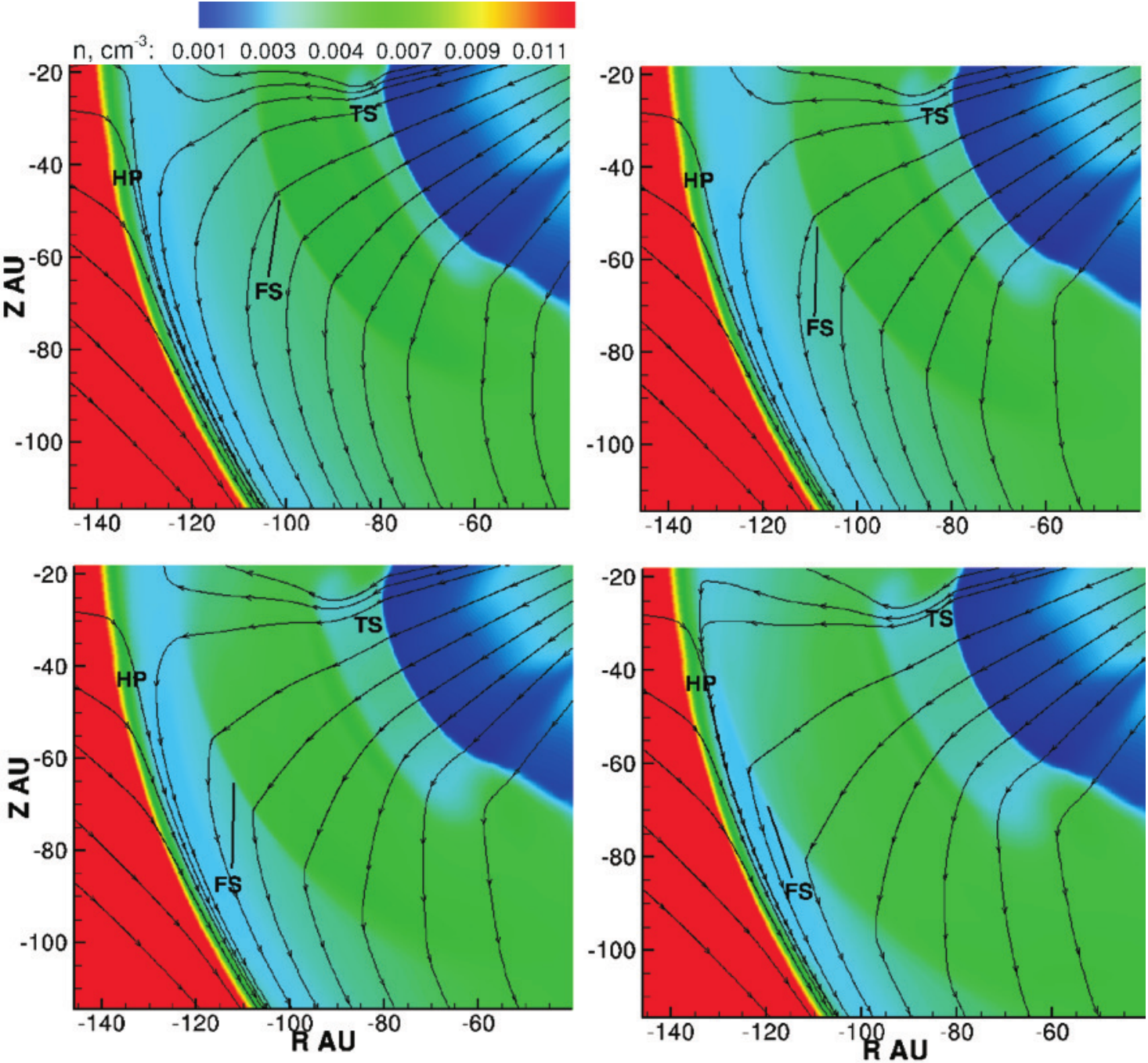}
\caption{Meridional cuts in the Oz-V2 plane from a 3D MHD simulation showing the distribution of the solar wind density ($cm^{-3}$) for different times when the interaction region propagates in the heliosheath. The plot (a) corresponds to the moment of time $t=1$yr; (b) $t=1.1$ yr; (c) $t=1.2$ yr; and (d) $t=1.3$ yr 
\label{fig6}}
\end{figure}	

\subsection{Propagation in the heliosheath}

The heliosheath is a region with spatial variations of the solar wind parameters that may affect the shock properties and amplitudes of the disturbances propagating from the supersonic solar wind. Along the Voyager 2 direction from the TS to the HP, the plasma decelerates and the magnetic field magnitude increases. A region with enhanced magnetic filed exists near the HP (see Fig. \ref{fig1}). In our simulation the forward shock transmitted from the supersonic solar wind into the heliosheath is a weak perpendicular shock. For a weak perpendicular shock a dependence of the shock speed on the background parameters is given by \citep{gurnett2005}
\[ V^2_{n}=2(c_{s}^2+c^2_{A})/(\delta - 1) (\delta_{max} - 1),
\]
where $V_{n}=\bf{(V_{sw}-D) \cdot n}$ is the normal component of upstream plasma velocity in the shock frame, $\bf{D}$ is the shock speed, $c_{s}$ is the sound speed upstream the shock, $c_{A}$ is the Alfven speed upstream the shock, $\delta$ is the ratio of upstream and downstream densities at the shock, and  $\delta_{max} = (\gamma+1)/(\gamma-1)$, where $\gamma=5/3$. In the heliosheath along Voyager 2, $c_{s}$ decreases and $c_{A}$ increases due to increase of magnetic field. The term $(c_{s}^2+c^2_{A})$ remains nearly constant across the heliosheath, which gives a constant $V^2_{n}$. But the solar wind velocity decreases across the heliosheath due to deceleration toward the HP. Therefore, a deceleration of the forward shock in the heliosheath should occur. Our simulation shows that the forward shock speed decreases by 10\%.  While the large-scale disturbance moves through the heliosheath, the compression of the shock also decreases by 6\%. The 3D effects of the flow become more pronounced when the forward shock approaches the HP. Figure \ref{fig6} shows plasma density and velocity streamlines in the V2-Oz plane in different moments of time while the disturbance propagates through the heliosheath. Near the HP $\delta_{FS}=1.3$ in Voyager 2 direction. Values of $\delta_{FS}$ within $\pm 10 \degr$ from the Voyager 2 direction show a latitudinal dependence: $\delta_{FS}$ increases by 5\% with increasing latitude and decreases by 10\%  from the Voyager 2 toward the equator. The forward shock becomes stronger with increasing latitude, because the solar wind turns around the heliopause to the heliospheric tails and compressed plasma behind the shock moves to the higher latitudes (Fig. \ref{fig6}). Our analysis of 3D effects was performed within the narrow highly resolved grid cone, but in general stronger 3D variations may exist in the heliosheath.

Our simulation shows an essential radial and latitudinal expansion of the interaction region in the heliosheath. After the TS crossing, the radial width of the interaction region is 22 AU; it increases by $60 \%$ closer to the HP. From Fig. \ref{fig6} it can be seen that the disturbed region significantly spreads in latitudinal extent. The interaction region with latitudinal extension  $\sim 30 \degr$ in the supersonic solar wind produces a large-scale disturbed region in the heliosheath with an extension of about $ 50 \degr$. 

\begin{figure}
\centering
\includegraphics[width=\hsize]{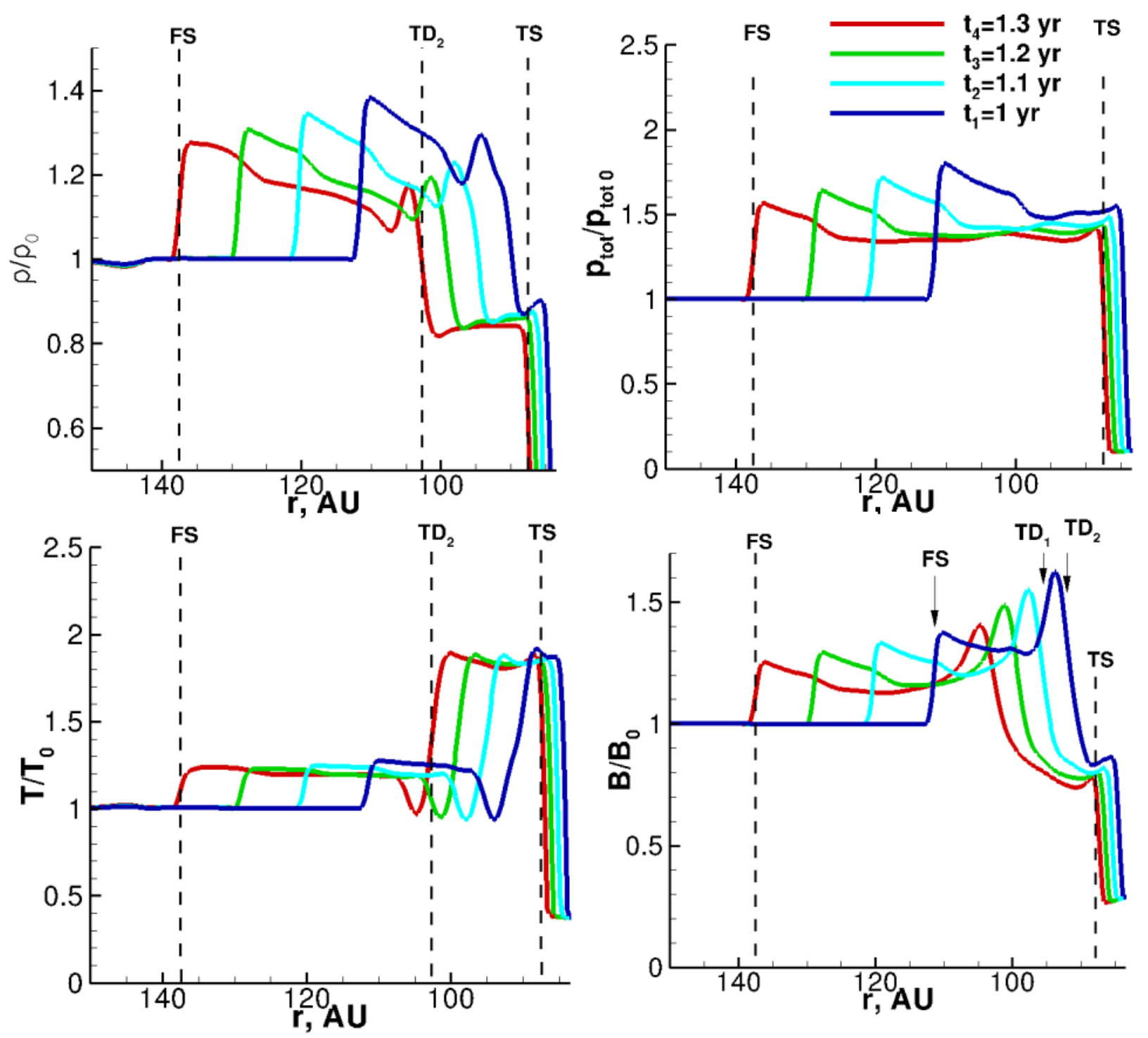}
\caption{ Profiles of the normalized density $\rho/\rho_0$, the total pressure $p_{tot}/p_{tot0}$, the temperature $T/T_0$,  and the  magnetic field $B/B_0$ along the Voyager 2 trajectory for different moments of time showing the evolution of the fluctuations when the interaction region propagates in the heliosheath. Notations: FS - forward shock, TS - termination shock, $TD_1$, $TD_2$ - tangential discontinuities.
\label{fig7}}
\end{figure}

Figure \ref{fig7} shows the normalized solar wind number density, the total pressure, the temperature, and the magnetic field along the Voyager 2 direction for different moments of time while a disturbance travels throughout the heliosheath. The disturbance is characterized by two peaks of the magnetic field: B increases at the forward shock FS and at the tangential discontinuity $TD_1$. At the forward shock the increase of $B/B_0$ is about $30 \%$, at the tangential discontinuity the magnetic field increases stronger, $ \sim 60 \%$. Density profiles show that the disturbance also creates two regions with maximum plasma density. The amplitudes of $\rho/\rho_0$ in both regions are comparable, $\sim 30 \% $. As the disturbed region moves toward the HP, the amplitude of $B/B_0$ fluctuation decreases by 12 \%, $\rho/\rho_0$ decreases by 11 \%, and $V/V_0$ increases by 20 \% along the Voyager 2 direction.  

\begin{figure}
\centering
\includegraphics[width=\hsize]{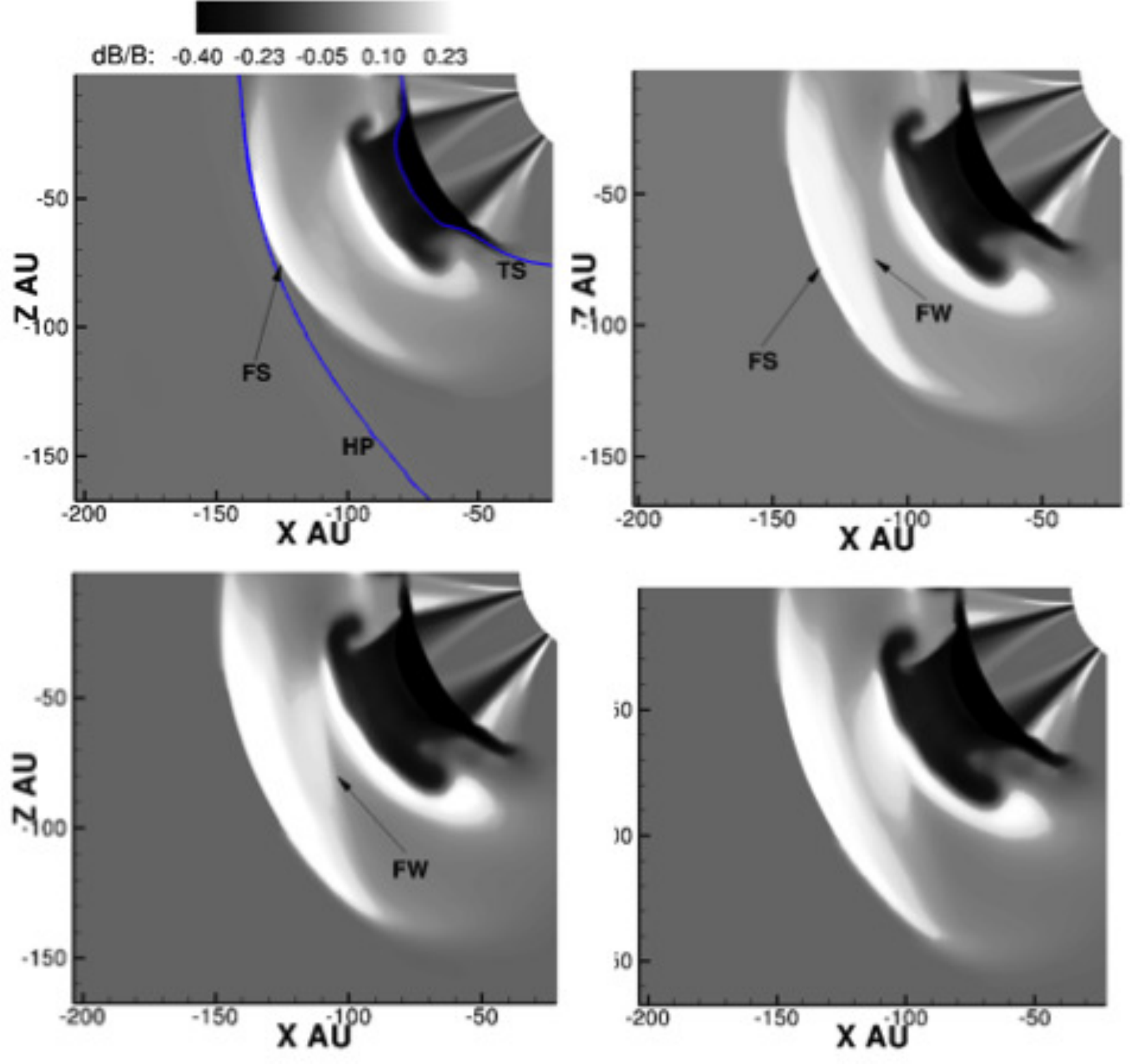}
\caption{Meridional cuts  through the the Oz-V2 plane from a 3D MHD simulation showing the contours of  $\delta B/B_0$ for different times. Notations: FS is the forward shock, HP is the heliopause, FW is reflected fast magnetosonic wave propagating toward the TS, and TS is the heliospheric termination shock. The temporal evolution is shown in an animation available in the online edition.  
\label{fig8}}
\end{figure}

One year after crossing the TS, the forward shock encounters the HP (left top plot in Figure 8). The interaction of the forward shock with the HP causes a motion of the HP away from the Sun by few AUs. As the result of the shock interaction with the HP, a new shock is created that propagates into the interstellar medium, a magnetosonic wave is reflected from the HP inside the heliosheath and propagates back toward the TS. Fig. \ref{fig8} presents the propagation of the reflected  wave in the heliosheath toward the TS. It shows contours of  $\delta B/B_0 = (B-B_0)/B_0$, where $B_0$ refers to the steady-state solution.
FS denotes the forward shock that propagates in the interstellar medium. FW denotes the magnetosonic wave reflected from the HP. The locations of TS and HP are shown in the left top plot in Fig. 8. The reflected magnetosonic wave FW is a fast magnetosonic wave. From our simulation, the average fast magnetosonic speed in the heliosheath is about 250 km/s. Using the results presented in Figure 8, one can calculate that the speed of reflected magnetosonic wave is about 230 km/s. This speed is comparable to the fast magnetosonic speed in the heliosheath. It can also be seen that vortices may form in the heliosheath at the sides of the high-speed stream. The formation of vortex flow and possible role of K-H instability will be discussed in subsequent paper.

 The reflected magnetosonic wave FW  approaches the TS. The interaction of the TS with FW occurs about 2.5 years after the forward shock entered the heliosheath. A bounce of the magnetosonic wave causes the displacement of the TS by 2-3 AU toward the Sun. Because of the complex flow in the heliosheath it is hard to identify  the secondary reflected waves from the TS. However, the formation of a cascade of small disturbances in solar wind plasma reflecting from the TS and propagating back to the heliosheath is seen from the simulation (see animation linked to Figure \ref{fig8} in the on-line material). After the interaction with the magnetosonic wave the TS moves into the heliosheath. This movement may also contribute to the formation of compressional waves in the solar wind plasma downstream of the TS, since the TS acts like a piston that pushes the solar wind material inside the heliosheath. These secondary magnetosonic waves reflected from the TS are caught up by the bulk solar wind flow around the HP and are carried away to the tails of the heliosphere.

\section{Conclusions}
Using a global 3D MHD model of the solar wind interaction with the LISM, we studied a propagation of a pair of forward and reverse shocks from the supersonic solar wind into the heliosheath. A pair of shocks is usually driven by MIRs that form in the solar wind during solar maximum.
The pair of shocks was initiated in the direction of the Voyager 2 trajectory. To capture newly created discontinuities formed in the solar wind through the passage of a pair of forward-reverse shocks, we used a non-uniform spatial grid with highest resolution of 0.5 AU in the cone along the Voyager 2 trajectory.

To study the propagation of the pair of shocks from the supersonic solar wind to the HP we analyzed (1) the evolution in the supersonic solar wind, (2) the interaction with the TS, (3) the propagation through the heliosheath, and (4) the interaction with the HP. The pair of shocks is formed by a steep increase  of the solar wind speed at 30 AU from 417 km/s to 625 km/s. The simulation in the supersonic solar wind showed that the interaction region between the shocks expands and decelerates; both shocks weaken while moving to the larger heliospheric distances. When the interaction region is at 45 AU from the Sun, the plasma density $\rho$, the temperature $T$, and the magnetic field strength $B$ vary inside the region by factors of 2.4, 2.5, and 2.3. While the interaction region propagates outward, the variation amplitudes of the plasma parameters change: in the Voyager 2 direction the density fluctuations $\rho/ \rho_0$ decrease by 12 \%, $T/T_0$ increases by 10 \%, $B/B_0$ increases by 6 \%. We found that magnetic field fluctuations behave differently depending on the latitude.

Modeling the propagation of a shock pair in the heliosheath showed the following effects: the collision of the pair of forward and reverse shocks with the TS causes the motion of the TS away from the Sun of a few AU and several new discontinuities are created downstream of the TS: a fast forward shock, tangential discontinuities, and possibly rarefaction waves and slow shocks. The passage of the forward shock through the TS results in a weakening of both shocks. The structure of the interaction region between the shocks changes after the TS crossing: in the heliosheath it is bounded by a weak forward shock at the front and a tangential discontinuity at the rear. Newly formed discontinuities in the heliosheath cause variations in the solar wind parameters and disturb the heliosheath flow. The magnetic field strength in the interaction region increases by 30-60\%, the density by 30\%, the temperature by 30\%. 

While the forward shock propagates toward the HP, the shock strength decreases and the shock decelerates.  Magnetic field and  density fluctuations have smaller amplitudes as the interaction region propagates deeper in the heliosheath. The interaction of the forward shock with the HP causes the outward motion of the HP and reflection of fast magnetosonic wave from the HP. Reflected waves propagate inside the heliosheath toward the TS and encounter the TS after about 1.5 year. The interaction of magnetosonic waves with the TS displaces the TS toward the Sun and generates a secondary reflection of the magnetosonic waves from the TS. The propagation of reflected waves between the TS and the HP contributes to the dynamic flow in the heliosheath solar wind.

In this study we considered an evolution of a shock pair that was initialized by an abrupt increase of the solar wind speed. In a subsequent study we will focus on modeling realistic solar events that produce a pair of shocks or one strong interplanetary shock in the heliosphere driven by MIR. For example, series of solar events during August-September 2005 detected at the Earth produced a MIR and an associated strong shock that were observed at Voyager 2 immediately before the TS (at 79 AU) and then at Voyager 1 beyond the TS. \citet{webber2007, webber2009} analyzed  temporal variations of $> 70 MeV$ cosmic ray intensities observed on Voyager 1 during the shock's travel through the TS and HP. Based on the data, they estimated the shock arrival times to the TS and the HP and the shock propagation speed in the heliosheath. A comparison of a realistic solar event simulation with observations would significantly improve our model and enable us to use it to predict heliosheath flows caused by the solar events during the present increase of solar activity.

\begin{acknowledgements}
We would like to thank the anonymous referee for his valuable comments. We also thank Astrid Peter for correction errors in English that improved the readability of this paper. This work was supported by National Science Foundation CAREER Grant ATM-0747654. The calculations were performed at NASA AMES Pleiades Supercomputer facilities. We acknowledge partial support of this work by Presidium RAS Program 22, RFBR grant 13-02-01467 and by The Ministry of education and science of Russian Federation (project 8413). The authors also acknowledge the International Space Science Institute (ISSI) in Bern, Switzerland. 
\end{acknowledgements}

\end{document}